# Projection-Based Reduced Order Model and Machine Learning Closure for Transient Simulations of High-Re Flows

My Ha DAO*, Hoang Huy NGUYEN

Institute of High Performance Computing, Agency for Science, Technology and Research,

1 Fusionopolis Way, #16-16 Connexis, Singapore 138632.

* Corresponding author, daomh@ihpc.a-star.edu.sg

ABSTRACT

The paper presents a Projection-Based Reduced-Order Model for simulations of high Reynolds turbulent flows. The PBROM are enhanced by incorporating various models of turbulent viscosity and residual closures to model the effects of interactions among the modes and energy dissipations. Remarkable improvements in prediction accuracies are achieved with a suitable turbulent viscosity model and a residual closure. The enhanced PBROM models are demonstrated for high-Re flows past a cylinder in two- and three- dimensions. These enhancements have shown capable of capturing complex flow features and removing unnecessary ones, while not affecting the efficiency of the overall model.

## 1 Introduction

Transient simulations of turbulent flows are of interest in many engineering applications. While the Direct Numerical Simulations (DNS) with almost all turbulence length-scales resolved are often too computationally expensive for an engineering application, the Large Eddy Simulation (LES) and Reynolds Averaged Navier-Stokes (RANS) simulations, in which some larger turbulence length scales are modelled, are often adopted. LES and RANS simulations are still computationally slow and are impractical for uses in applications that required near-real-time predictions such as cyber-numerical simulations, model predictive controls, etc. In those applications, surrogate models such as those based on Artificial Neural Networks (ANN) or Reduced-Order Model (ROM) are often the optimal choices. In this work, we focus on the ROMs that are built from the pre-simulated data and the governing equations, called projection-based reduced order models.

Projection-based reduced order models (PBROM) are projections of high-fidelity numerical models onto a reduced subspace and is widely used for modelling fluid dynamical systems [1, 2, 3]. PBROM model is stable and accurate for modelling of low-Re flows with a suitable number of basis vectors [4, 5, 6]. For simulations of realistic turbulent flow, important considerations in PBROM are the model of turbulent viscosity and the truncation of low energy modes.

The former has similar effect as in the high-fidelity models. This is particularly true for high-Re flow as the turbulent viscosity is strongly dependent on the velocity fields. Even high-fidelity models often require a closure to compute the turbulent viscosity which directly affects the dissipation of energy of the whole system collectively. Several attempts to incorporate turbulent viscosity in PBROM models have been carried out, from a global constant coefficient for all modes [7] to a linear viscosity kernel that increases the dissipation linearly with the modal index [8] and a complex viscosity model that is temporal and modal dependent [9].

The latter relates to the losses of flow information and the nonlinearity (interaction of modes) due to the truncation of low energy modes, inheriting from the construction of PBROM models.

The effects of discarded low energy modes in high-Re flows are complex. The most direct effect is that PBROM simulations of these flows often require more modes in order to capture important features. That leads to significant increase of PBROM model size. However, even a complete set of modes is not sufficient to reproduce the full order dynamics accurately. In nonlinear dynamical systems, the low energy modes (or high-order modes) and the interactions of these modes with the high energy modes in PBROM simulations could be responsible for energy transfer among the modes and energy dissipation, which is analogous to the concept of large eddy simulations [10]. Energy transfers between modes in ROM modelling of a turbulent flow were studied [11]. It is observed that energy transfers are local in the POD basis with a net forward energy cascade. Energy transfers between the mean flow and successive modes as well as transfer of energy between the modes were also observed in [12] based on and proper orthogonal decomposition (POD) on a PIV data of a medium Re flow. Hence, it is important to model the effect of energy transfer in PBROM. This effect is physically not accountable in the turbulent viscosity model.

Recent works treat the two effects together by introducing a regularization term to the governing equation to account for the effects of both the turbulent viscosity and the truncated modes [13, 14, 15, 16, 17]. The stabilization parameter in the form of artificial viscosity that is estimated dynamically at each time step by a supervised neural network approach [14] or a least-squares based estimation [15]. San and Maulik [16] proposed an Extreme Machine Learning closure model to correct the temporal evolution of the PBROM's momentum equations to account for literally all lost information in the construction of PBROM modes (as compared to the original high-fidelity models) including the turbulent viscosity, the discarded modes and the interactions of these modes with other modes. The method was applied to a 2D Bousinessq equations. Another similar work applied on a simpler 1D Burgers equation [17] but with a different type of network, the Deep Residual Network [18].

We attempt to model the turbulent viscosity and the closure for discarded modes separately by several methods and test against various cases governed by fully nonlinear Navier-Stokes equations with different complexities. Our findings on a 2D test cases show that a suitable viscosity model can produce a right amount of energy dissipation and keep the amplitude of the time-evolution of the modal coefficient stable. However, the frequency of time-evolution of the modal coefficient is shifting and some super-harmonics appear. The frequency shift and the growth of super-harmonics reduce significantly when a NN closure model is introduced. The turbulent viscosity and NN closure models also work for a complex 3D flow but with less improvement as compared to the 2D case. In our work, the Extreme Machine Learning (ELM) model [16, 19] and the Nonlinear Autoregressive Exogenous (NARX) model [20, 21, 22] are employed.

In this paper, we will present a brief on the construction of a PBROM model based on OpenFOAM platform, details of the models for turbulent viscosity and NN closures, test cases on high-Re flows in 2D and 3D and the results, and finally the discussions and conclusions.

## 2 PBROM-ML model

### 2.1 Projection-based ROM (PBROM)

The equations governing incompressible turbulent flows comprise of a set of partial differential equations (PDEs) which, in Cartesian coordinates $\mathbf{x} = (x, y, z)$, have the form of

$$\mathbf{u}_t = -\mathbf{u} \cdot \nabla \mathbf{u} + \nu_E \Delta \mathbf{u} - \nabla p / \rho \qquad (1)$$
$$\nabla \cdot \mathbf{u} = 0 \qquad (2)$$

where the velocity $\mathbf{u} = (u, v, w)$, the pressure $p$ are spatio-temporal functions in a computational domain $\Omega_f$; $\rho$ is the density of the fluid. The effective viscosity $\nu_E$ comprises of the molecular viscosity constant $\nu_M$ and the turbulent kinematic viscosity $\nu_T$ (or turbulent viscosity, in short) which varies in time and space, $\nu_E = \nu_M + \nu_T$.

The governing PDEs can be solved by many available Computational Fluid Dynamics (CFD) packages such as Fluent, StarCCM (commercial) or OpenFOAM (open-source). In this study, the "pisoFoam" solver in OpenFOAM [23] is employed. The solver uses the Finite Volume Method (FVM) for spatial discretisation and the Pressure Implicit with Splitting of Operator (PISO) algorithm to solve the discretised NS equations [24, 25].

A PBROM can be constructed from the PDEs and a solution subspace, $S^\omega$. As shown in [4, 5], a variable $\omega$ could be decomposed in the flowing form

$$\omega = \omega_0 + \mathbf{\Phi}^\omega a^\omega + \mathbf{G}^\omega b^\omega \tag{3}$$

where $\mathbf{\Phi}^\omega := \{\phi_i^\omega\}_{i=1}^{r^\omega} \in \mathbb{R}^{N \times r^\omega}$ and $\mathbf{G}^\omega := \{g_i^\omega\}_{i=1}^{q^\omega} \in \mathbb{R}^{1 \times q^\omega}$ are the matrices comprising of the solution subspace for the inner domain, $\phi_i^\omega$, and the boundary condition functions for the boundary, $g_i^\omega$, respectively; $N$ is the number of unknowns of a variable of the CFD model; $a^\omega := \{a_i^\omega\}_{i=1}^{r^\omega}$ are modal coefficients, and $a_i^\omega = (\omega, \phi_i^\omega) \in \mathbb{R}$; $b^\omega := \{b_i^\omega\}_{i=1}^{q^\omega}$ are the boundary condition values; $r^\omega$ is the number of basis vectors and $q^\omega$ is the number of boundaries.

The PDEs with variables $\omega := u, v, w, p$ are projected on the respective solution subspaces spanning $r^\omega$ basis vectors, $S^\omega := \mathbf{\Phi}^\omega$, in an off-line phase to obtains the projection-based reduced-order model (PBROM) which comprise of a set of ordinary differential equations (ODEs) corresponding to the PDEs as

$$a_t^\mathbf{u} = f_\mathbf{u}(a^\mathbf{u}, a^p) + g_\mathbf{u}(b^\mathbf{u}, b^p) \tag{4}$$
$$a^p = f_p(a^\mathbf{u}) + g_p(b^\mathbf{u}, b^p) \tag{5}$$

where $a^\mathbf{u} := [a^u, a^v, a^w]^T$ and $a^p$ are the modal coefficients; $b^\mathbf{u} := [b^u, b^v, b^w]^T$ and $b^p$ are the boundary conditions for velocity and pressure, respectively.

Important considerations in PBROM simulations of turbulent flows are the model of turbulent viscosity and the truncation of low energy modes. The former has direct effect to the simulation through $\nu_E$ whilst the latter relates to the flow information and the nonlinearity (interaction of modes) lost due to the truncation of low energy modes.

### 2.1.1 Models for turbulent viscosity

In this study, several models for turbulent viscosity based on the ensemble of turbulent viscosity fields collected from the CFD simulation are considered for the PBROM. The first model is the temporal-spatial average of the ensemble, $\overline{\langle \nu_T \rangle}$, as shown in (6). The second model is the temporal-average, $\langle \nu_T \rangle$, in (7). The third model comprises of temporal-average, $\langle \nu_T \rangle$, and the 1st POD mode of the turbulent viscosity, $a_1^{\nu_T} \phi_1^{\nu_T}$, as shown in (8).

$$\nu_E = \nu_M + \overline{\langle \nu_T \rangle} \tag{6}$$
$$\nu_E = \nu_M + \langle \nu_T \rangle \tag{7}$$
$$\nu_E = \nu_M + \langle \nu_T \rangle + a_1^{\nu_T} \phi_1^{\nu_T} \tag{8}$$

The first model is the simplest as $\overline{\langle \nu_T \rangle}$ is a constant scalar, hence no modification to the projection (4), (5) is needed. In the second model, $\langle \nu_T \rangle$ is a constant vector representing the mean field of turbulent viscosity. An additional projection for $\langle \nu_T \rangle \Delta \mathbf{u}$ is required for the construction of the PBROM. The third model considers the time-varying component of the turbulent viscosity, $a_1^{\nu_T} \phi_1^{\nu_T}$. While the same projection for $\langle \nu_T \rangle \Delta \mathbf{u}$ can be applied for $\phi_1^{\nu_T} \Delta \mathbf{u}$ as $\phi_1^{\nu_T}$ is also a constant vector, the coefficient $a_1^{\nu_T}$ needs to be predicted online. A full projection of a turbulence closure or a machine learning model could be used for modelling $a_1^{\nu_T}$. Model (8) can be extended to include more modes for the turbulent viscosity. As the focus of this study is on the effects of $\nu_E$ in PBROM simulations, we employed a simple spline interpolation for the prediction of $a_1^{\nu_T}$.

*2.1.2 Closure for discarded low energy modes*

PBROM model was shown stable and accurate for modelling of low-Re flows with a suitable number of basis vectors [4, 5]. The effects of discarded low energy modes in high-Re flows are more complicated. The most direct effect is that these flows often require more modes in order to capture important turbulence features. That leads to significant increase of PBROM model size. More importantly, the low energy modes, or high-order modes, and the interactions of these modes with the high energy modes in PBROM simulations may be responsible for energy transfer among the modes and energy dissipation, which is analogous to the concept of large eddy simulations [26].

To account for the residual effects of the discarded modes in the momentum equations, San and Maulik [16] proposed an NN closure model to correct the temporal evolution of the PBROM. In simulations of turbulent flows, the NN closure is expected to account for the information loses and the interactions with other modes. The momentum equation (4) can be rewritten as

$$a_t^{\mathbf{u}} = \mathfrak{R}^{\mathbf{u}} + \widetilde{\mathfrak{R}}^{\mathbf{u}} \tag{9}$$

where, $\mathfrak{R}^{\mathbf{u}} = f_p(a^{\mathbf{u}}) + g_p(b^{\mathbf{u}}, b^p)$; $\widetilde{\mathfrak{R}}^{\mathbf{u}}$ is the residual that is modelled by an NN closure. During the PBROM simulations, the projection of truncated modes, $\mathfrak{R}^{\mathbf{u}}$, is first evaluated following (4). The residual $\widetilde{\mathfrak{R}}^{\mathbf{u}}$ is closed by the NN model with $\mathfrak{R}^{\mathbf{u}}$ being the input, $\widetilde{\mathfrak{R}}^{\mathbf{u}} = f^{NN}(\mathfrak{R}^{\mathbf{u}})$. Figure 1 shows two possible NN architectures for a $a_t^{\omega}$ equation, where $\omega \equiv u, v, w$. In this work, the Extreme Learning Machine (ELM) and Nonlinear AutoRegressive eXogenous (NARX) methods are used to train the function $f^{NN}$. The networks are briefly described below while more details can be found in literature [16, 19, 20, 21, 22].

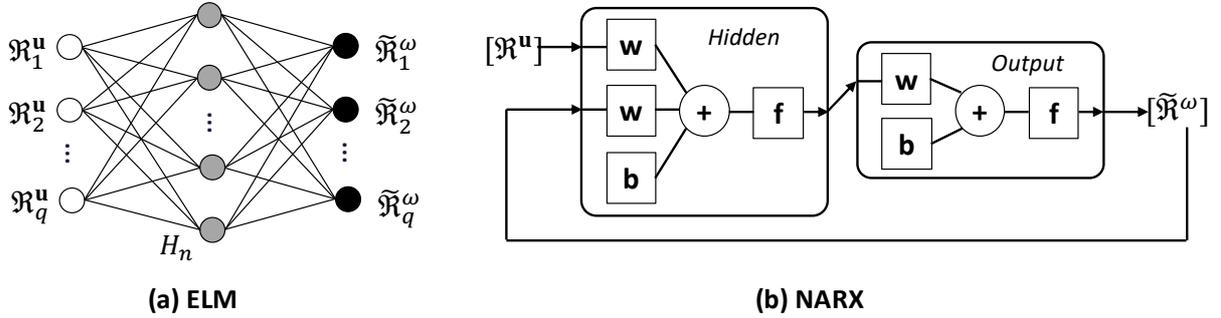

(a) ELM  (b) NARX

Figure 1 ELM and NARX networks used for residual closures for PBROM momentum equations

## 2.2 Extreme Learning Machine (ELM)

The ELM is a single layer feed-forward neural network that is based on the least square approximations (Figure 1a). The ELM training leads to the optimal weight $\mathbf{W}_{opt}^2$ which is used to construct the mapping function $f^{NN}$ for prediction of the residual $\widetilde{\mathfrak{R}}^{\mathbf{u}}$ during the PBROM simulation. The mapping function can be presented in a compact form,

$$f^{NN}(\mathfrak{R}^{\mathbf{u}}) := \mathbf{W}_{opt}^2 G(\mathbf{W}^1 \mathfrak{R}^{\mathbf{u}} + \mathbf{B}^1) \tag{10}$$

where $\mathbf{W}^1$ and $\mathbf{B}^1$ are weight and bias matrices of non-zero random numbers associated with the numbers of neurons in the hidden layer; $G(\mathbf{X})$ is a tan-sigmoid function acting on a matrix $\mathbf{X}$. The optimal weight is computed during the training of the network as $\mathbf{W}_{opt}^2 = \mathbf{H}^{\dagger}[\widetilde{\mathfrak{R}}^{\mathbf{u}}]_S^T$. Here, the residual $[\widetilde{\mathfrak{R}}^{\mathbf{u}}]_S$ is the target matrix of the training; $\mathbf{H}^{\dagger}$ is a pseudo-inverse of a matrix $\mathbf{H}$ representing the hidden layer neurons, where $\mathbf{H}^T = G(\mathbf{W}^1 [\mathfrak{R}^\mathbf{u}]_S + \mathbf{B}^1)$; the script $^T$ indicates transpose operator. In these test cases, the number of hidden layer neurons set at 10.

The residuals $[\widetilde{\mathfrak{R}^\mathbf{u}}]_S$ correspond to the collected snapshots $\mathbf{u}$ and subspace $S$ of selected basis vectors $\mathbf{\Phi}$ and are computed by applying the projection of the PDEs given in (1) on the subspace and subtracting the projection of truncated modes the same subspace, $[\widetilde{\mathfrak{R}^\mathbf{u}}]_S = (\mathbf{u}_t, \mathbf{\Phi}^\mathbf{u}) - [\mathfrak{R}^\mathbf{u}]_S$. The projection of truncated modes, $[\mathfrak{R}^\mathbf{u}]_S$, whose columns correspond to the columns of the target matrix, is the input of the training.

### 2.3 Nonlinear AutoRegressive Exogenous Model (NARX)

The close-loop Nonlinear AutoRegressive eXogenous (NARX) architecture is also used to predict the residual $\widetilde{\mathfrak{R}^\mathbf{u}}$ from $\mathfrak{R}^\mathbf{u}$. In a NARX neural network, the prediction is related to the past outputs and both the current and past inputs via two parameters of feedback delay and input delay. The data structure of inputs and outputs of the NARX model is described in Figure 1b. In this application, the NARX toolbox in Matlab 2018b [27] is employed. Both delay parameters are set to one. The Levenberg-Marquardt backpropagation algorithm is selected for the training. The network parameters are as follows: number of hidden layers is 10, processFcns = 'mapminmax', min_grad = 1e-15, divideFcn = 'divideblock' and divideParam = 0.8, 0.15, 0.05 for the train, validation and test blocks, respectively. Other parameters are set to default. The projection of truncated modes, $[\mathfrak{R}^\mathbf{u}]_S$, and the residuals, $[\widetilde{\mathfrak{R}^\mathbf{u}}]_S$, as described in the ELM section above, are the input and the output of the training.

Unlike the ELM, the trained NARX networks, $f^{NN}(\mathfrak{R}^\mathbf{u})$, are in the forms of Matlab's network structures. After training, these networks are stored and will be loaded and called during the PBROM simulations (also in the same Matlab environment).

## 3 Simulation results and discussions

In this section, the PBROM with ELM and NARX models as residual closures are demonstrated for turbulent flows past a 2D stationary circular cylinder and a 3D square column. The full CFD simulations to collect snapshots are carried out in OpenFOAM. The ensemble means, basis vectors (or POD modes) and modal coefficients (or POD coefficients) for velocities and pressure are computed using POD method. The projections of the differential operators on the basis vectors are done using OpenFOAM platform. The PBROM solution algorithms are implemented in Matlab. Details of the POD, projections and solution algorithms are described in authors' previous work [4, 5].

### 3.1 Definitions of test cases and PBROM models

#### 3.1.1 Flow past a 2D circular cylinder

Figure 2 show configuration of flows past a two-dimensional stationary circular cylinder having diameter of $D = 1$ m. A constant velocity, $\mathbf{u}_{in} = (u_{in}, v_{in}) = (1, 0)$ ms$^{-1}$, is imposed at $\Gamma_{in}$. Zero gradient for velocity, $\nabla \mathbf{u} = 0$, is applied at $\Gamma_{top} \cup \Gamma_{bottom} \cup \Gamma_{out}$. Non-slip velocity, $\mathbf{u} = 0$, is applied on the cylinder surface. Zero gradient for pressure, $\nabla p = 0$, is imposed at all boundaries except for $\Gamma_{out}$ where $p = 0$. The fluid density is $\rho = 1$ kgm$^{-3}$. The molecular viscosity is chosen as $\nu_M = 10^{-4}$ m$^2$s$^{-1}$ and $10^{-6}$ m$^2$s$^{-1}$, corresponding to Reynolds numbers of $Re = uD/\nu_M = 10^4$ and $10^6$. A total of 124,338 computational cells are used to discretize the computational domain for the full CFD simulation.

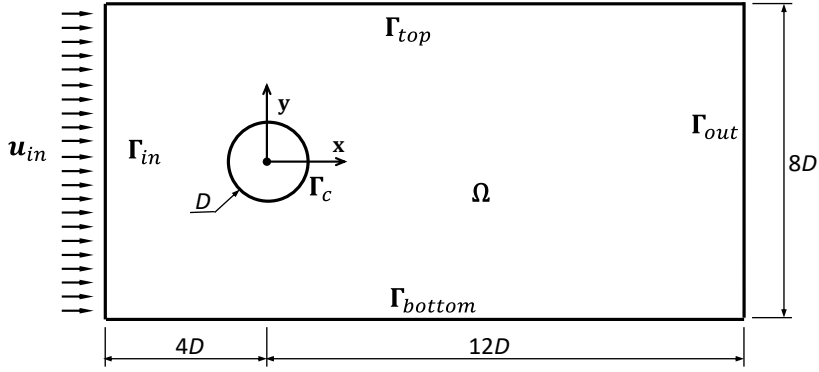

Figure 2 Schematic description of the flow past a 2D stationary cylinder problem.

The full CFD simulations use the "pisoFoam" solver and the $k$-$\omega$-SST turbulence closure [28]. The simulations are initialized from a steady stage solution, i.e. the flow field with a symmetric recirculation bubble behind the cylinder [29]. As the flow develops, the recirculation bubble becomes unstable, and vortices start to grow and shed downstream. After the flow field has reached a quasi-steady state, snapshots of velocity and pressure are collected at an interval of $\delta t_E = 0.1$ s over a period of $T_E = 30$ s. That gives 301 snapshots for each variable. Figure 3 shows snapshots of fully developed velocity magnitude and pressure at an arbitrary time instance for the case of $Re = 10^6$.

The forces acting on the cylinder comprise of normal pressure force, $\mathbf{F}_p = \rho \sum_i \mathbf{s}_{f,i}(p_i - p_{ref})$ and tangential viscous force, $\mathbf{F}_v = \rho \sum_i \mathbf{s}_{f,i} \cdot (\nu_E \mathbf{R}_i)$, where $\mathbf{s}_{f,i}$ is the face area vector at and $\mathbf{R}_i$ is the deviatoric stress tensor of the velocity at the boundary face $i$ of the cylinder. The forces from PBROM and CFD reculsts are computed following the same OpenFOAM's function for a consistency in the comparison.

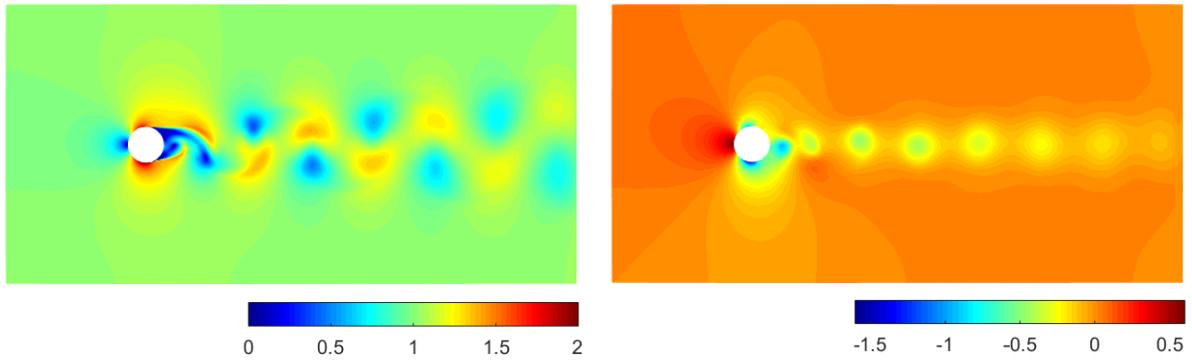

Figure 3. Normalized velocity magnitude and pressure obtained from the CFD simulation of 2D flow, $Re = 10^6$.

### 3.1.2 Flow past a 3D square cylinder

Figure 4 shows the schematic description of flow past a 3D stationary square cylinder with $D = 3.85$ mm. The settings at boundaries are similar to that of the 2D cylinder case. The inlet velocity is $\mathbf{u}_{in} = (u_{in}, v_{in}, w_{in}) = (1, 0, 0)$ ms$^{-1}$, giving $Re = 3850$. The computational domain is discretized into approximately 4.2 million cells.

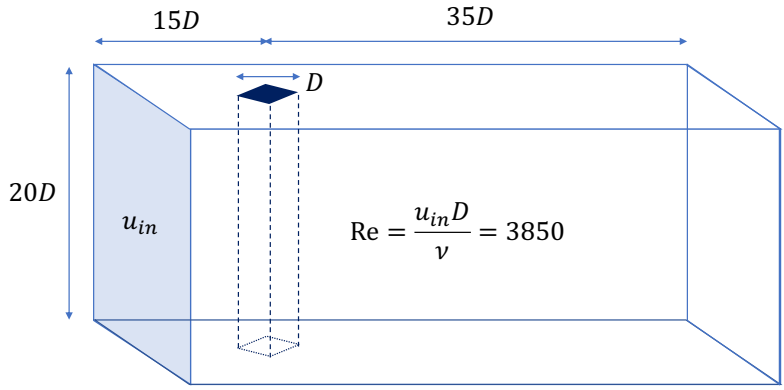

Figure 4. Schematic description of the flow past a 3D square column

The CFD simulation uses the "pisoFoam" solver and $k$-$\omega$-SST-SAS closure [30]. A set of 170 snapshots of flow fields are collected after the flow has stabilized for the construction of PBROM model. That covers approximately 10 cycles of primary vortex shedding. Figure 5 shows the complex 3D structures of a fully developed velocity field obtained from the CFD simulation at an arbitrary time instance.

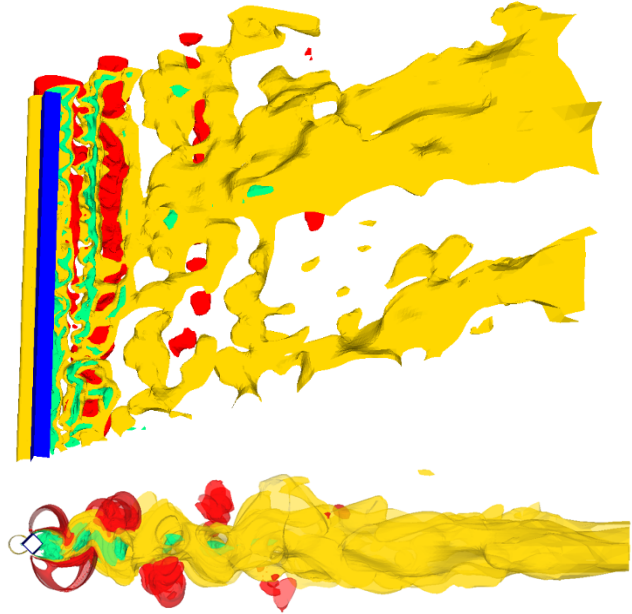

Figure 5. Snapshot of fully developed velocity contours from CFD simulation of flow past a 3D column. Top: side-view at center plane; Bottom: top-view at center plane. Color demacates velocity magnitudes.

### 3.1.3 PBROM models

Various PBROM models with different turbulent viscosity models and with/without residual closures for discarded modes are studied. The models are summarized in Table 1.

Table 1 PBROM models with different turbulent viscosity models and closures.

| Model | Turbulence model | Closure |
|---|---|---|
| A  | $\nu_E = \nu_M + \overline{\langle \nu_T \rangle}$ |  |
| A1 | $\nu_E = \nu_M + \overline{\langle \nu_T \rangle}$ | $\widetilde{\mathfrak{R}}^{\text{ELM}}$ |
| B  | $\nu_E = \nu_M + \langle \nu_T \rangle$ |  |
| C  | $\nu_E = \nu_M + \langle \nu_T \rangle + a_1^{\nu_T} \phi_1^{\nu_T}$ |  |
| D | $v_E = v_M + \langle v_T \rangle$ | $\widetilde{\mathfrak{R}}^{ELM}$ |
| E | $v_E = v_M + \langle v_T \rangle + a_1^{v_T} \phi_1^{v_T}$ | $\widetilde{\mathfrak{R}}^{ELM}$ |
| F | $v_E = v_M + \langle v_T \rangle$ | $\widetilde{\mathfrak{R}}^{NARX}$ |

## 3.2 A sensitivity study on number of basis vectors

Unlike linear problems, where errors due to discarded modes are expected to decrease as number of modes used for constructions of the PBROM models, nonlinear interactions among modes may lead to changes in the behaviours of the error curves. Hence, convergence studies are conducted to choose the numbers of modes used for constructions of the PBROM models. In this study the RMSE of POD coefficients for velocities and pressure are compared. The RMSE is computed for every mode $a_i^\omega$ as $RMSE = \sqrt{1/n_T \sum_{j=1}^{n_T}[(a_{ij}^\omega)_P - (a_{ij}^\omega)_E]^2}$, where $(a_{ij}^\omega)_P$ and $(a_{ij}^\omega)_E$ are the predicted (PBROM) and exact (CFD) solutions of mode $a_i^\omega$ at time $t_j$ and $n_T$ is the number of data points. The PBROM model A is used for the case of flow past 2D cylinder at Re = $10^4$, PBROM model B is used for the cases of flow past 2D cylinder at Re = $10^6$ and flow past 3D square column.

Figure 6 shows the RMSEs of POD coefficients obtained from PBROM models constructed from different numbers of POD modes for the three test cases. Results show good convergences of the PBROM models for all cases. The running time of these models are very fast, about $\mathcal{O}(10^4)$ times faster than that of the full CFD simulations. Based on the study, numbers of POD modes chosen for the PBROM models in the subsequent studies are: $r^u = r^p = 10$ for the case of flow past 2D cylinder at Re = $10^4$, $r^u = r^p = 20$ for the case of flow past 2D cylinder at Re = $10^6$, $r^u = r^p = 60$ for the case of flow past 3D square column.

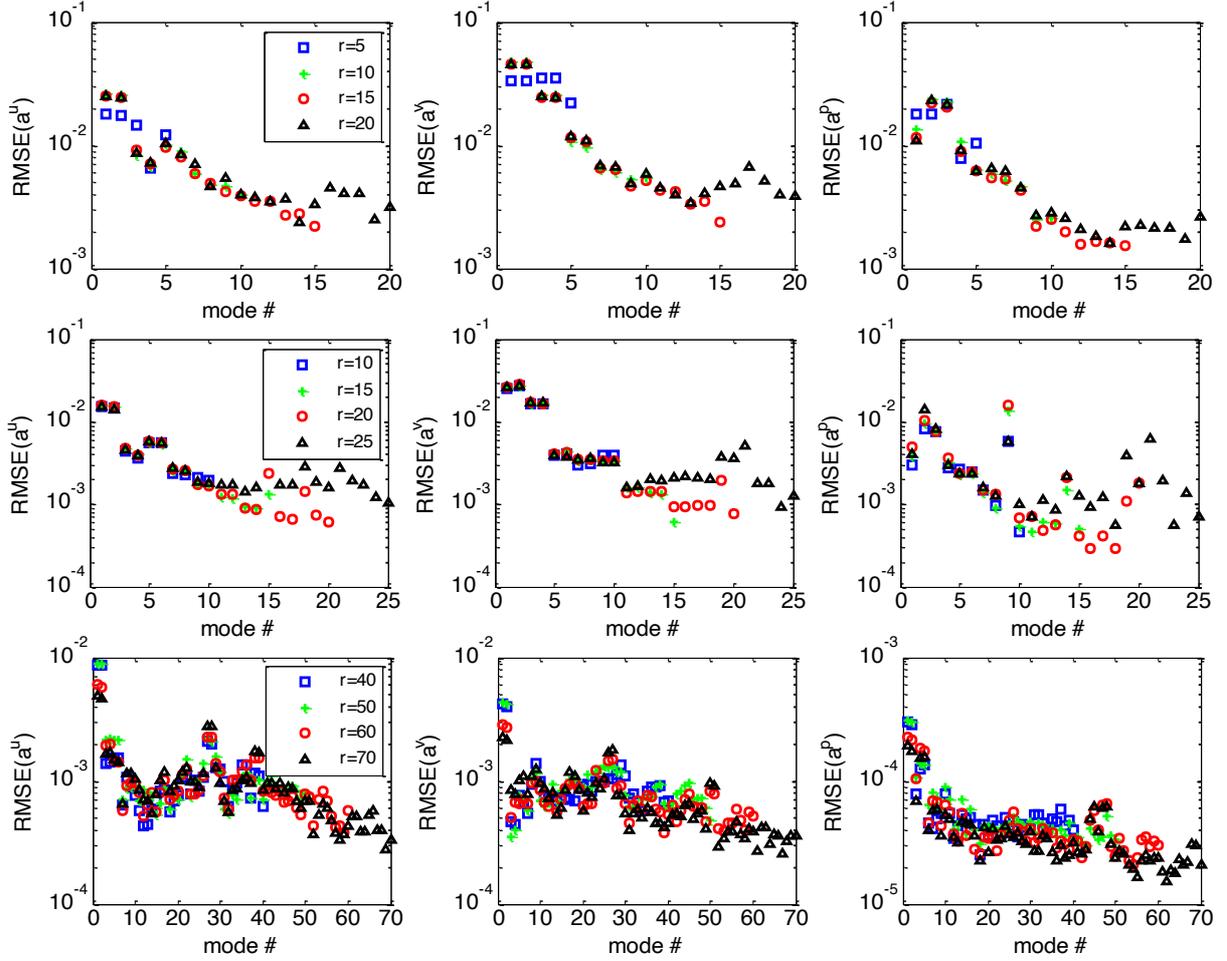

Figure 6. RMSE of POD coefficients for velocities and pressure predicted by PBROM construtcuted from different numbers of basis vectors. Top row: for 2D cylinder, Re = $10^4$; Middle row: for 2D cylinder, Re = $10^6$; Bottom row: for 3D cylinder.

### 3.3 PBROM simulation results for 2D flow, Re = $10^4$

Shown in Figure 7, the simplest PBROM model A with a constant temporal-spatial averaged turbulent viscosity, $\overline{\langle \nu_T \rangle}$, is able to capture the dynamics of the three highest energy modes in the simulation of $Re = 10^4$ flow. There is a noticeable decreasing of the modal amplitudes which can be numerically reduced by slightly tuning the turbulent viscosity constant in model A. On the other hand, with the ELM residual closure, i.e. model A1, the decreasing of modal amplitudes is completely removed without adjusting the turbulent viscosity constant.

It is demonstrated that the PBROM alone with a tuned turbulent viscosity constant work well for cases with simple flow structures. However as presented in the next sections, with more complex flow fields (higher Re or three-dimensional), the simple turbulent viscosity model does not work well. In those cases, the averaged field of turbulent viscosity $\langle \nu_T \rangle$ and residual closures may be needed.

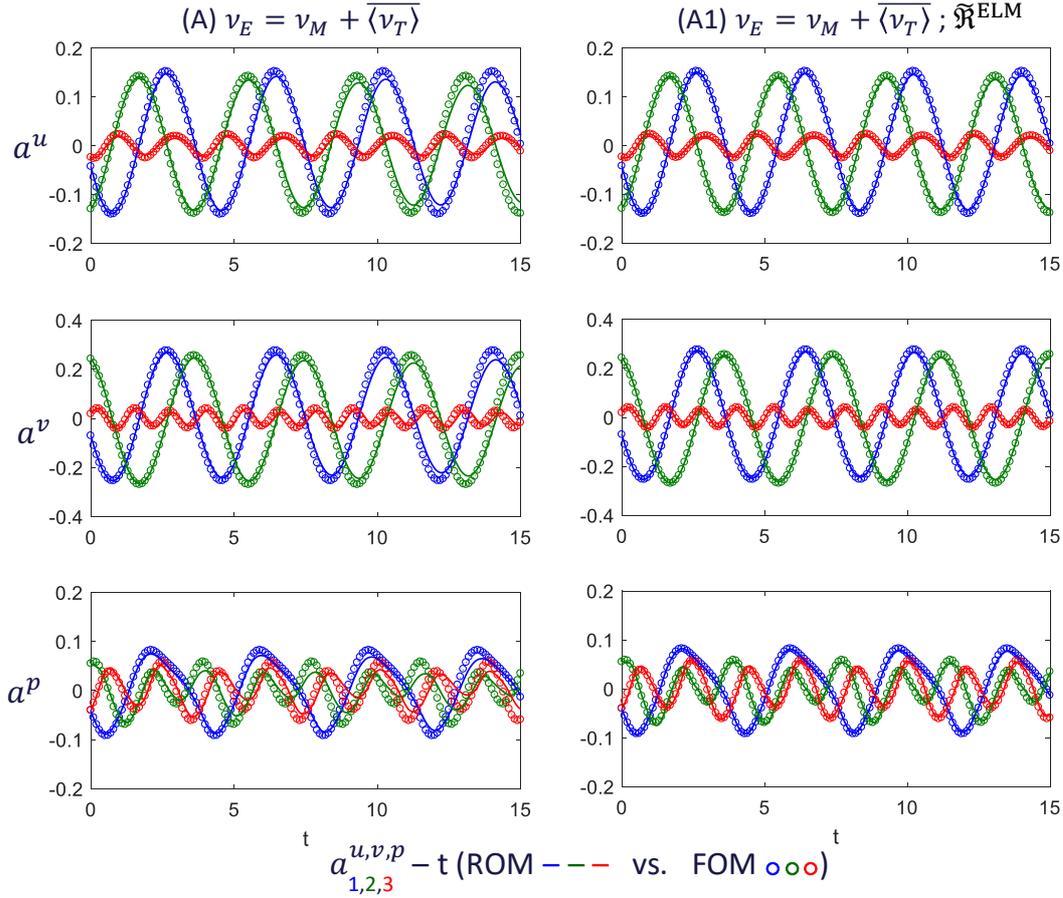

Figure 7. Time-series of modal coefficients for velocity and pressure, $a^{u,v,p}$, obtained from PBROM (solid lines) and CFD (empty circles) simulations for $Re = 10^4$. Colour code: blue – mode 1, green – mode 2, red – mode 3.

### 3.4 PBROM simulation results for 2D flow, Re = $10^6$

In this section, the PBROM models A, B, C, D, E, F are used for simulations of the $Re = 10^6$ flows. Figure 8 – Figure 10 show the time-series of modal coefficients from PBROM and CFD simulations. Figure 11 show the comparisons for modal coefficients of a higher order mode for velocities (mode 7[th]). The RMSE of POD coefficients for velocities and pressure are shown in Figure 12. Figure 13 – Figure 14 show the differences in the velocity and pressure fields between the PBROM and CFD simulations. Figure 15 plots the forces acting on the cylinder surface.

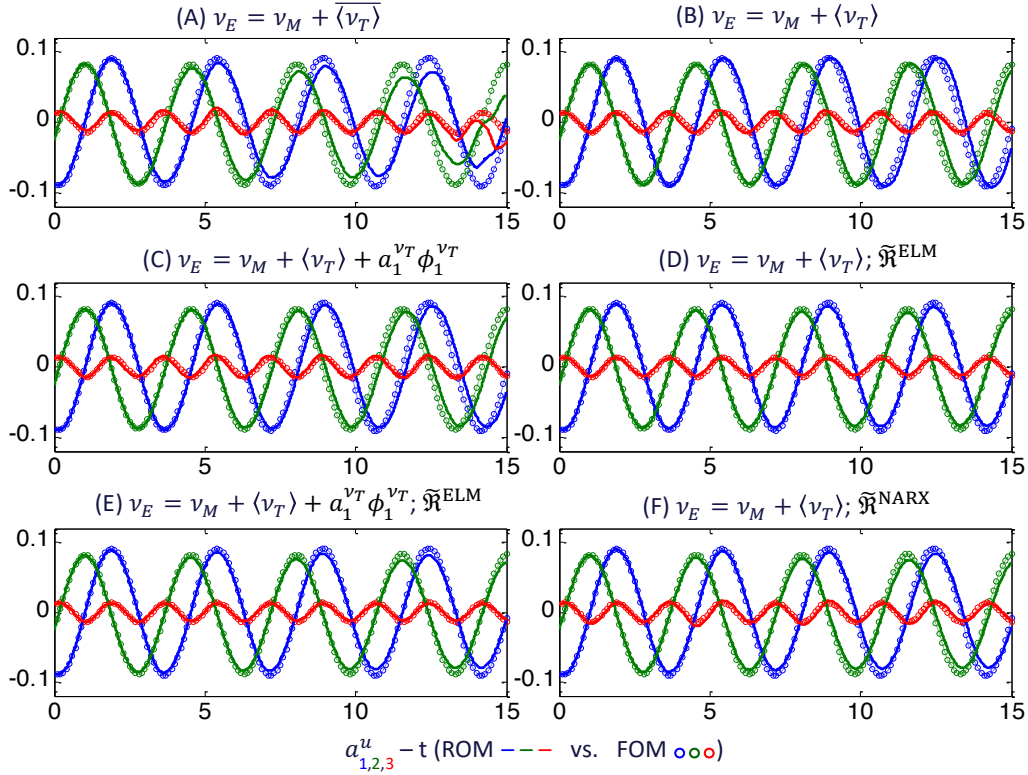

Figure 8. Time-series of modal coefficients for $u$-velocity, $a^u$, obtained from PBROM (solid lines) and CFD (empty circles) simulations. Colour code: blue – mode 1, green – mode 2, red – mode 3.

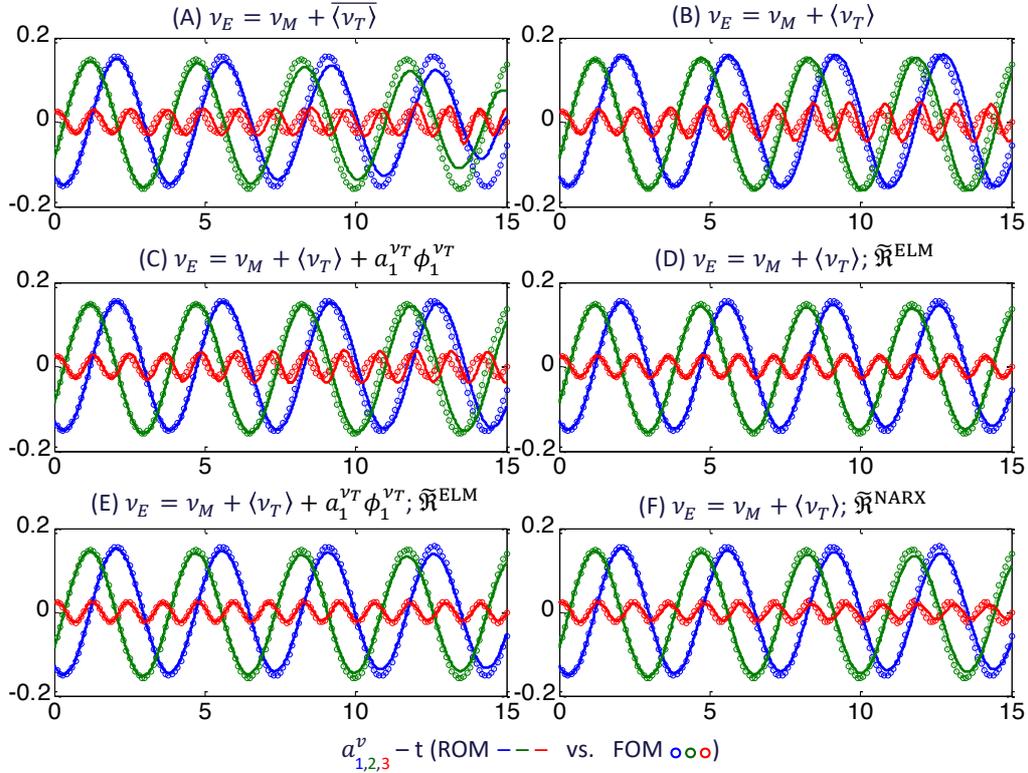

Figure 9. Time-series of modal coefficients for $v$-velocity, $a^v$, obtained from PBROM (solid lines) and CFD (empty circles) simulations. Colour code: blue – mode 1, green – mode 2, red – mode 3.

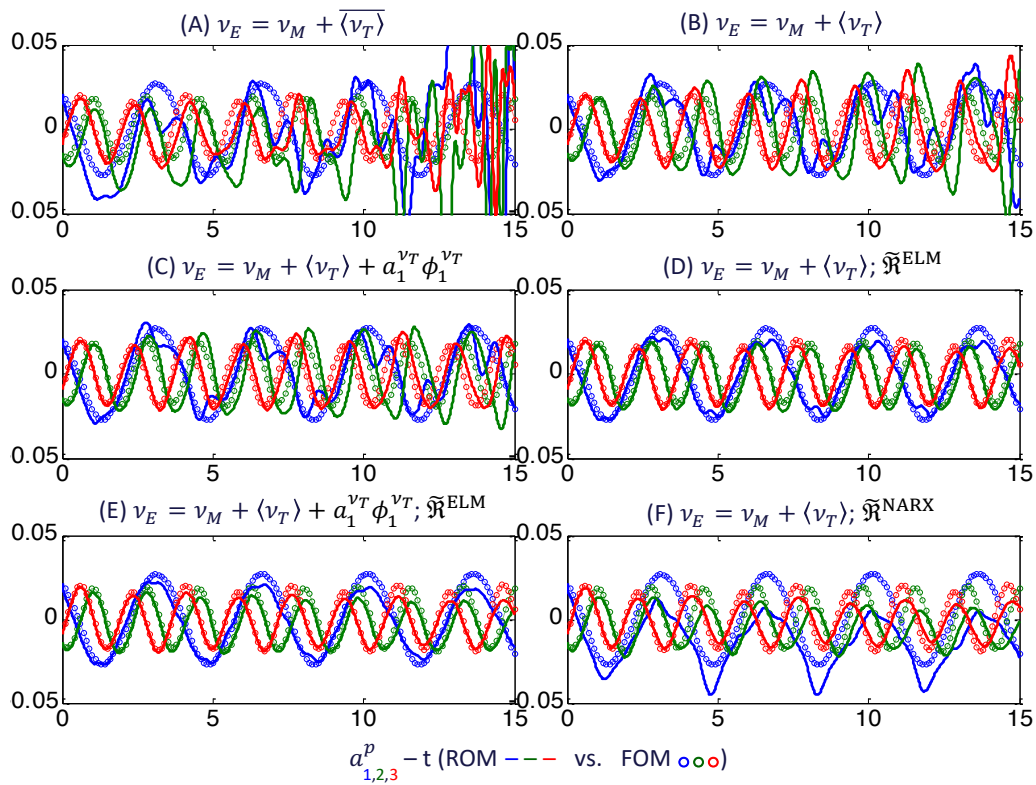

Figure 10. Time-series of modal coefficients for pressure, $a^p$, obtained from PBROM (solid lines) and CFD (empty circles) simulations. Colour code: blue – mode 1, green – mode 2, red – mode 3.

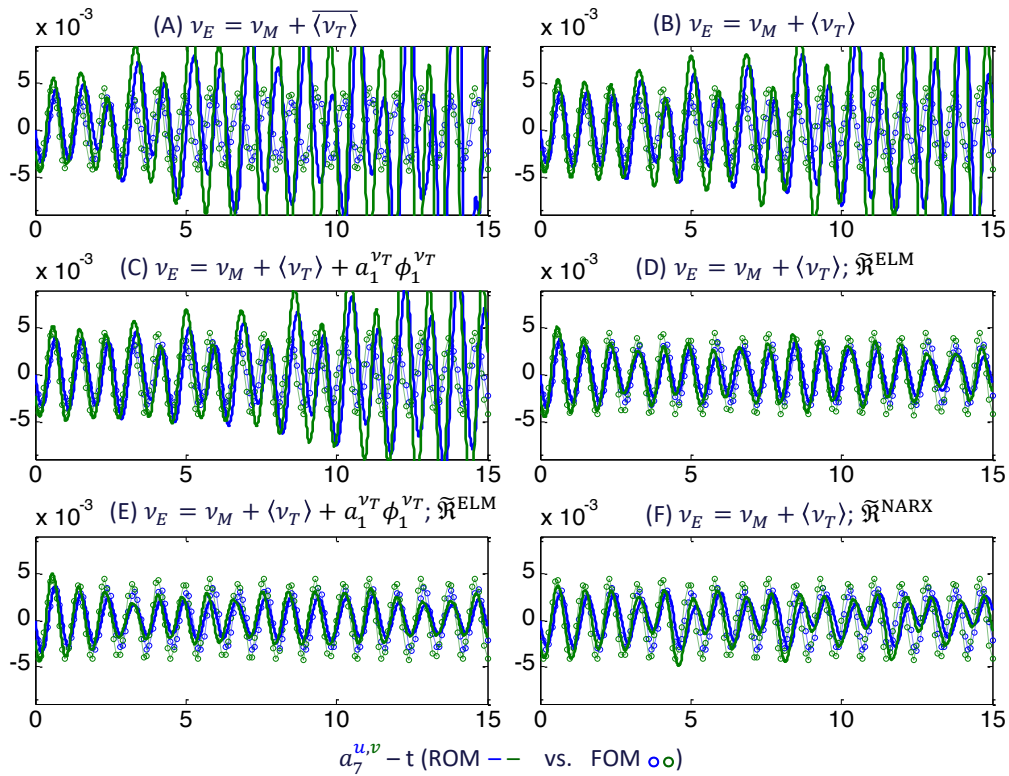

Figure 11. Time-series of modal coefficients for velocities for mode 7$^{\text{th}}$ obtained from PBROM (solid lines) and CFD (empty circles) simulations. Colour code: blue – $a^u$, green – $a^v$.

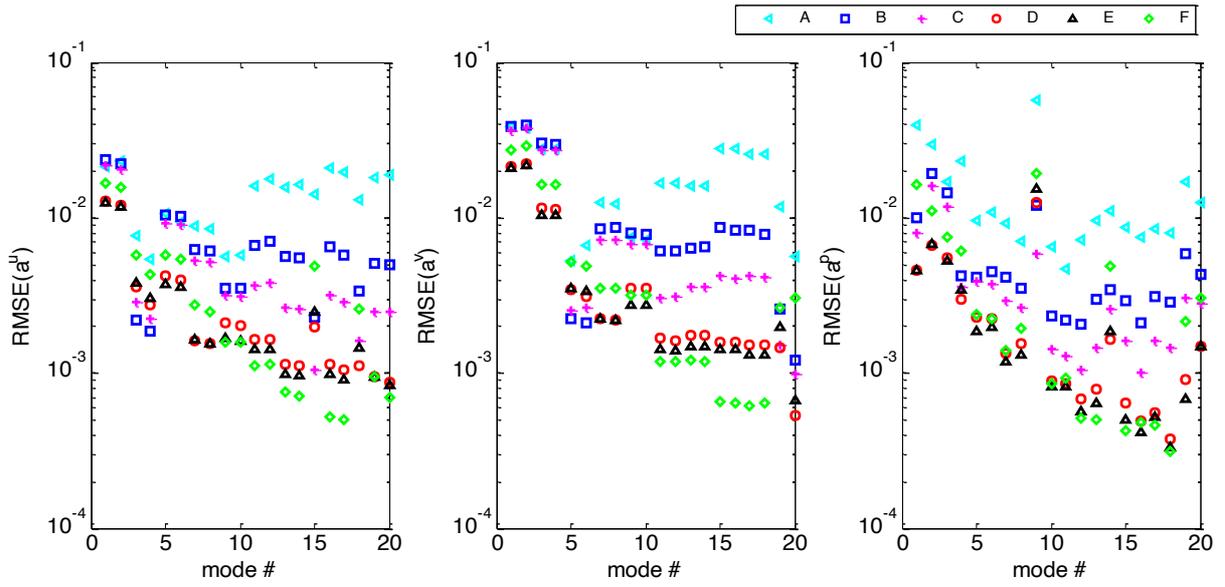

Figure 12. RMSE of POD coefficients for velocities and pressure predicted by different PBROM models.

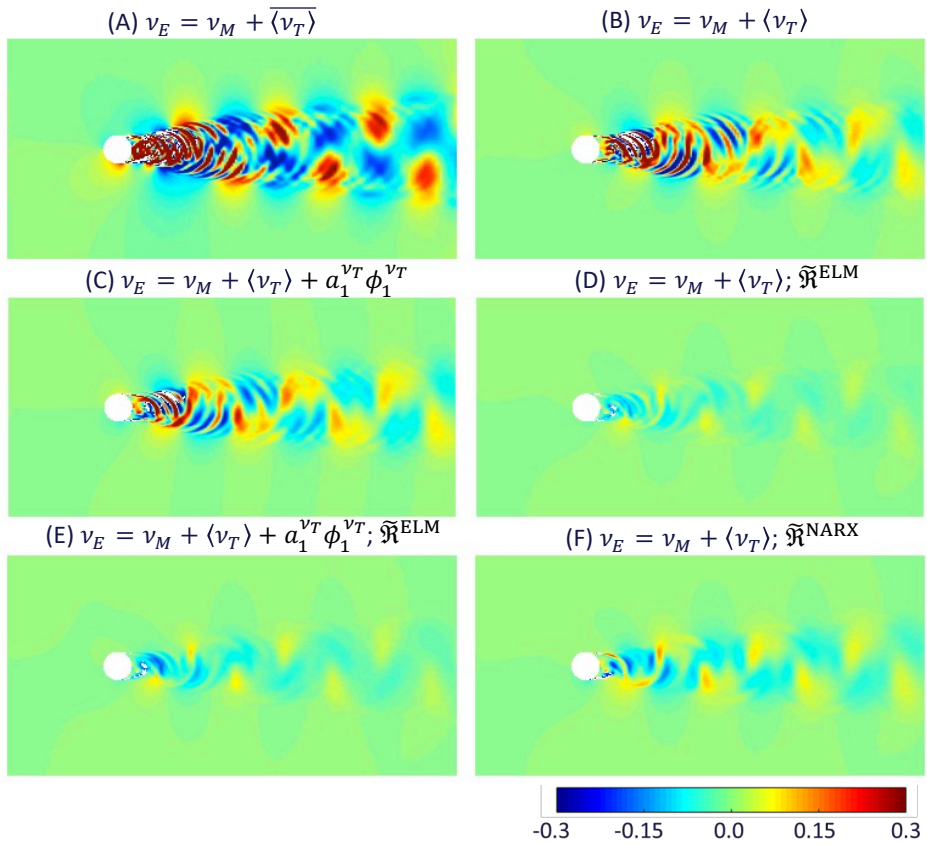

Figure 13. Deviations of normalized velocity between PBROM and CFD simulations at $t = 15\ s$.

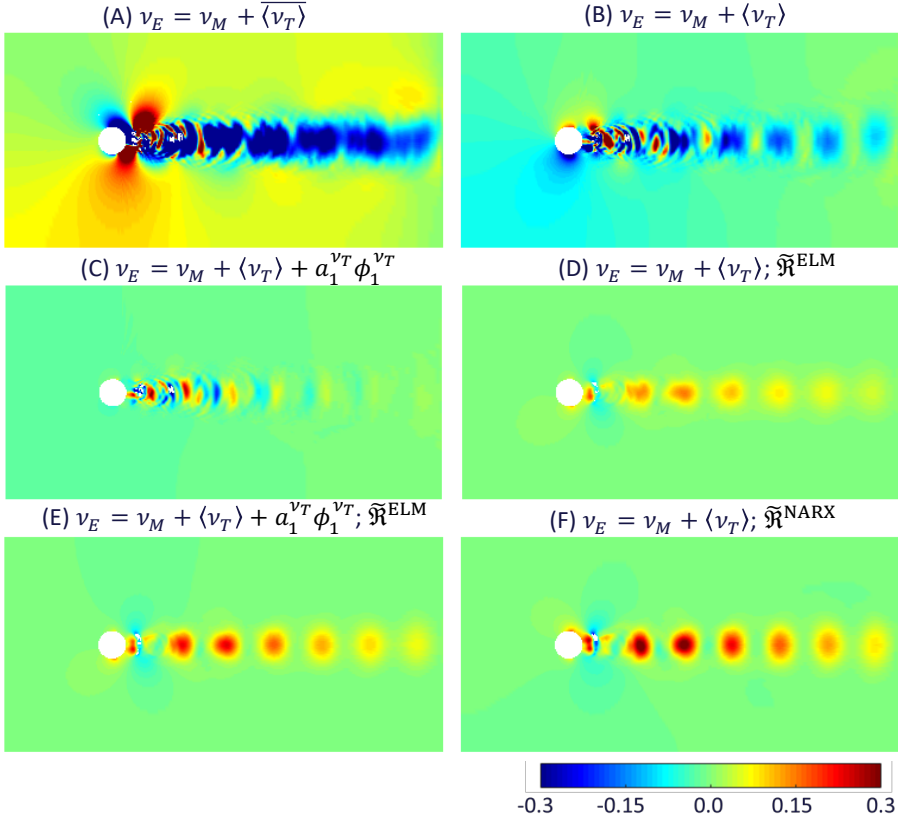

Figure 14. Deviations of normalized pressure between PBROM and CFD simulations at $t = 15\ s$.

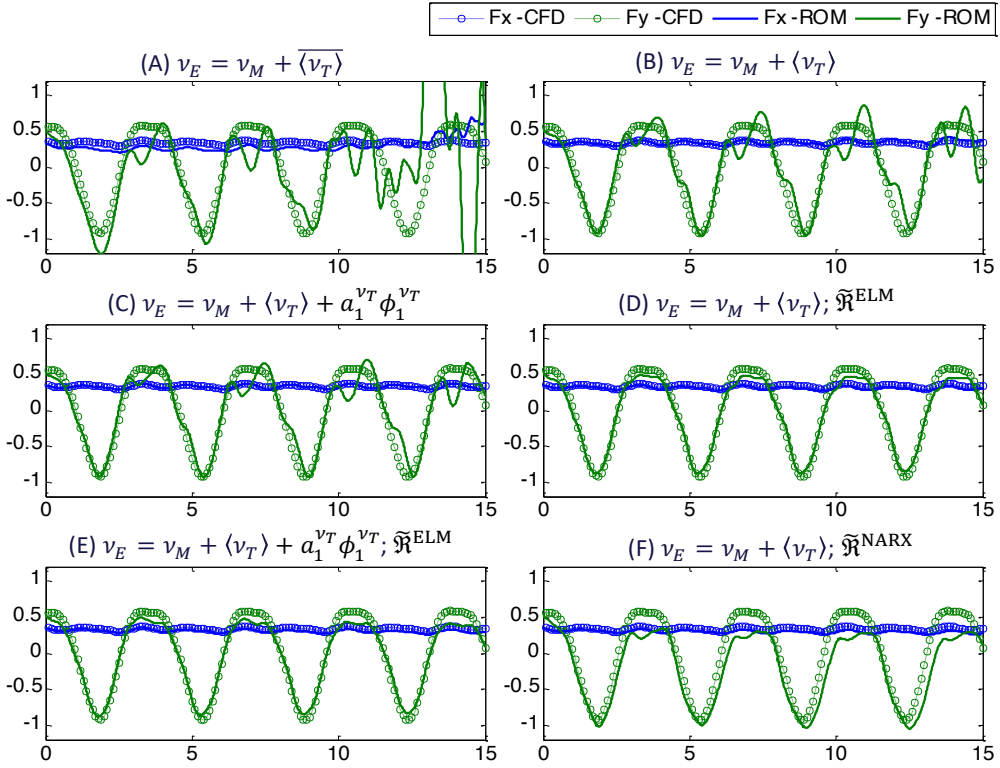

Figure 15. Time-series of the forces acting on the cylinder obtained from PBROM (solid lines) and CFD (symbolled lines) simulations.

### 3.4.1 Effect of turbulence viscosity model

Changes in the models for turbulent kinematic viscosity, i.e. models A, B, C in the order of increasing of complexity in Figure 8 – Figure 10, show significant improvements in prediction quality. With the temporal-spatial averaged turbulent viscosity ensemble (model A), the modal coefficients of velocity gradually deviate from the CFD results. In particular, magnitudes of mode 1 and 2 are decreasing. Super-hamonic oscillations are obserevd in the pressure coeffcient plots (Figure 10) for all three models. Modal coefficients of pressure become unstable near the end of the simulation period of model A. With the spatial variation of the turbulent viscosity (models B) the PBROM predictions become stable and more accurate. The amplitudes of the velocity coefficients maintain close to the CFD solution. With the temporal variation of the turbulent viscosity (model C), the prediction of velocity coefficients is slightly better such as in mode #3, while that of pressure coefficient is significantly improved although there still some super-harmonic oscillations occur. The differences are clearer in the results of a lower energy modes of velocity (see Figure 11 A, B, C). The turbulence viscosity models do affect significantly the evolutions of low energy modes of velocity, and in turn affect the evolution of pressure modes. The plots RMSE of the coefficients in Figure 12 show significant improvements from models A to C for mode #3, 4, 7 and higher modes.

Figure 13 – Figure 14 A, B and C show the improvements due to increasing of the complexity of the turbulence viscosity model reflected in the flow fields. As shown at time $t = 15s$ (approximately 3-4 vortex cycles), model B performs much better than model A. With the temporal variation of the turbulent viscosity, $a_1^{v_T}\phi_1^{v_T}$, model C predicts both velocity and pressure fields significantly better than model B, especially in the region near the cylinder surface. However, the deviations of model C from the CFD are still as much as 15% for velocity (0.3 over 2.0 in normalized terms) and 20% for pressure (0.3 over 1.5 in normalized terms).

The calculated forces on cylinder surface show similar trend with models B and C being more accurate than model A. However, there still exist super harmonics on the transverse force component in the results of both models B and C, which can be linked to the predictions of coefficient for pressure.

Noting that, in this simulation, the coefficient $a_1^{v_T}$ calculated using a spline interpolation from the known values. In a true prediction, a full projection of a turbulence closure or a machine learning model are needed for modelling $a_1^{v_T}$. That will increase the complexity and the susceptibility to errors of the PBROM model significantly while the gain in the prediction of flow fields is not significant.

### 3.4.2 Effect of ML closures

The ELM-based closure for the discarded modes in the momentum equations is added to models B and C, resulting in models D and E, respectively. The NARX closure is added to model B creating model F.

As one can see in Figure 8 and Figure 9, the modal coefficients for velocity obtained from PBROM models E and F simulations match very closely with the exact solution for the first highest energy modes. Significant improvements can be seen in the results of modal coefficients for $v$-velocity and pressure. One can notice that there are small but growing frequency shifts in the results of velocity mode 1 and mode 2 of models A, B, C regardless of turbulent viscosity models used. These frequency shifts are almost removed in the results of model D and E. Hence, this effect could be attributed to the contributions of discarded low energy modes to the high energy modes which are recovered by the closure. Furthermore, models D and E are also able to produce very good results for the low energy mode (Figure 11 D, E).

Remarkable improvements due to incorporating the closure can be seen in the predictions of modal coefficients for pressure (Figure 10 D, E). PBROM results match very well with the CFD results and the super-harmonic oscillations are almost removed. Comparing between the results from two models, model E shows better performance in mode 1 in removing the super-harmonic, while model D result is significantly more accurate in mode 2 and 3. Using the NARX-based closure (model F), improvements in predictions of modal coefficients for velocity compared to the PBROM model without closure (model B) can be achieved for both low and high energy modes. However, the improvement is less as compared to that achieved by the ELM-based closure (model D). Obvious differences can be observed in the results of modal coefficients for pressure. Compared to model B, model F is stable and produce much less super-harmonic oscillation, but the trajectory of mode 1 deviates away from the exact one. Compared to model D, model F is less accurate in this case. The plots RMSE of modal coefficients in Figure 12 show improvements up to two orders at low energy modes (mode #5 and above) in model D, E and F. These remarkable improvements help to improve the stability and accuracy in the time-evolution of the PBROM models.

In the form of reconstructed flow fields, simulations of PBROM with models D and E show similar low level of errors with respect to the CFD results at time $t = 15s$ (Figure 13), though model D result looks slightly better. Errors are around 10% for velocity (0.15 over 2.0 in normalized terms) and pressure (0.15 over 1.5 in normalized terms). Model F produces higher errors, but is better than model B in overall, especially in regions around the cylinder which is important for the prediction of forces acting on the cylinder. As one can observed from Figure 15, the turbulent viscosity models and the closures help to improve the accuracy of force prediction.

In overall, model D with the spatial variation of the turbulent viscosity and the ELM closure shows the best performance. While the turbulent viscosity models can help to stablize the model, achieve correct amplitudes of modal coeffcients in a long run, the closures help to improve in capturing frequency, removing high-frequency oscillation that probaly arise from nonlinear interactions and energy transfers among modes.

### 3.5 PBROM simulation results for 3D flow, Re = 3850

In this test case, the flow structure is complex, and the domain of simulations is much larger due to the expansion into a three-dimension space (see Figure 5). At Re = 3850, the vortex structures have developed into the third dimension along the column [31]. The interaction among modes and turbulent energy transferring happen in all three dimensions. In order to capture all flow features, 60 modes are required for each variable. The PBROM models B, D and F are constructed with the temporal-averaged $\langle v_T \rangle$ model for turbulent viscosity and the two ML closures.

Figure 16 show the time-series plots of the first three modes of velocity and pressure from the three PBROM simulations and derived from the CFD results. The flow is obviously dominant by the horiontal velocity, i.e. the $u$ and $v$ components, and PBROM models B, D are able to capture the dynamics of these two velocity components. The differences are at the $w$-component and the pressure. Without the closure, the magnitude of the 2$^{nd}$ and 3$^{rd}$ modes of the $w$-component keep growing. With the ELM closures (model D), the $w$-component and the pressure modes follow very well the trends of the CFD. With the NARX closure (model F), the results of the $w$-component is slightly better. Results of the other two components and pressure are good initially but become worse after approximately 3 cycles ($t \cong 5s$). There are significant shifts in the frequencies of all three modes as compared to the CFD results, but no obvious improvement in frequency shift is observed when closures are incorporated.

Figure 17 shows the RMSE of modal coefficients for velocities and pressure predicted by the three PBROM models over the first 5s. The RMSE plots show a mixed response for the $u$-component. The RMSEs of the $v, w$-components and pressure are smaller for almost all modes in the results of model D and F.

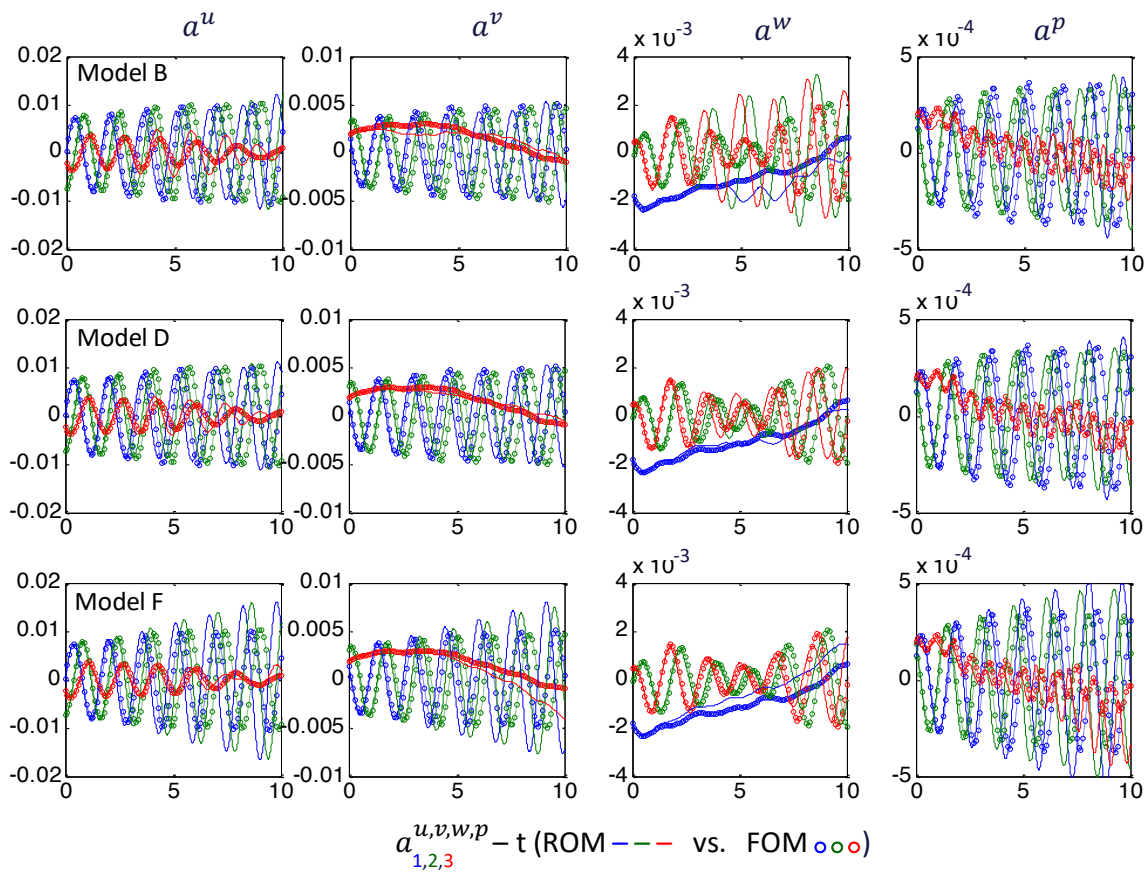

Figure 16. Time-series of the first 3 modes of velocity and pressure from PBROM models (solid lines) and derived from CFD results (symbolled lines); blue – mode 1, green – mode 2, red – mode 3.

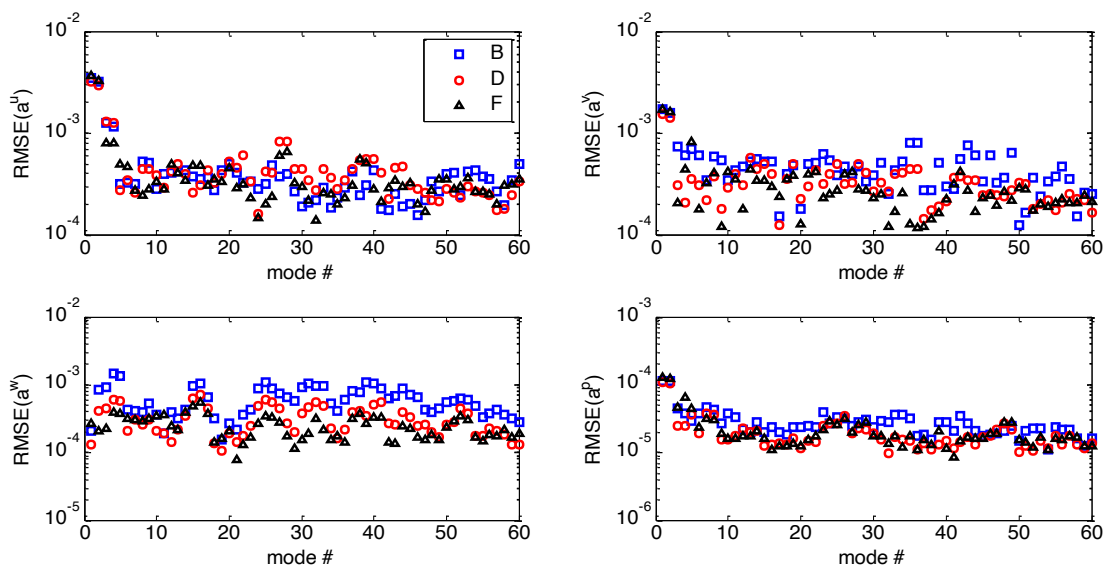

Figure 17. RMSE of POD coefficients for velocities and pressure predicted by different PBROM models.

The velocity fields from CFD, PBROM modes are extracted on a vertical and a horizontal centre planes for comparison and calculation of the deviations between the PBROM models and the CFD. These velocity fields and deviations are normalized to the inlet velocity. Results at $t = 5s$ (approximately 3-4 vortex shedding cycles) are shown in Figure 18. All three PBROM model results are quite resemble to that of CFD including vertical flow features. Zooming down to the region downstream of the lower end of the column in the vertical plane on the left, the PBROM model B predicts slightly stronger velocity magnitude of the shedding vortices, indicating by darker red spots while the flow structure near the top end of the column is picked up well. The PBROM model D captures the vertical distribution of the vortices slightly better. Model F captures the flow structures much better than the other two models. On the horizontal plane, the vortex structures behind the column obtained from model F is the most resemble to the CFD result.

Velocity differences from CFD results on the two centre planes are almost at the same magnitudes as shown in Figure 19. Spots of high error mainly associate with the high velocity spots. Magnitudes of the error at these spots are at approximately 30%. These errors could be explained by the frequency shifts observed in the time-series plots of the modal coeffcients of velocity and pressure in Figure 16. At the area around the cylinder, errors are small.

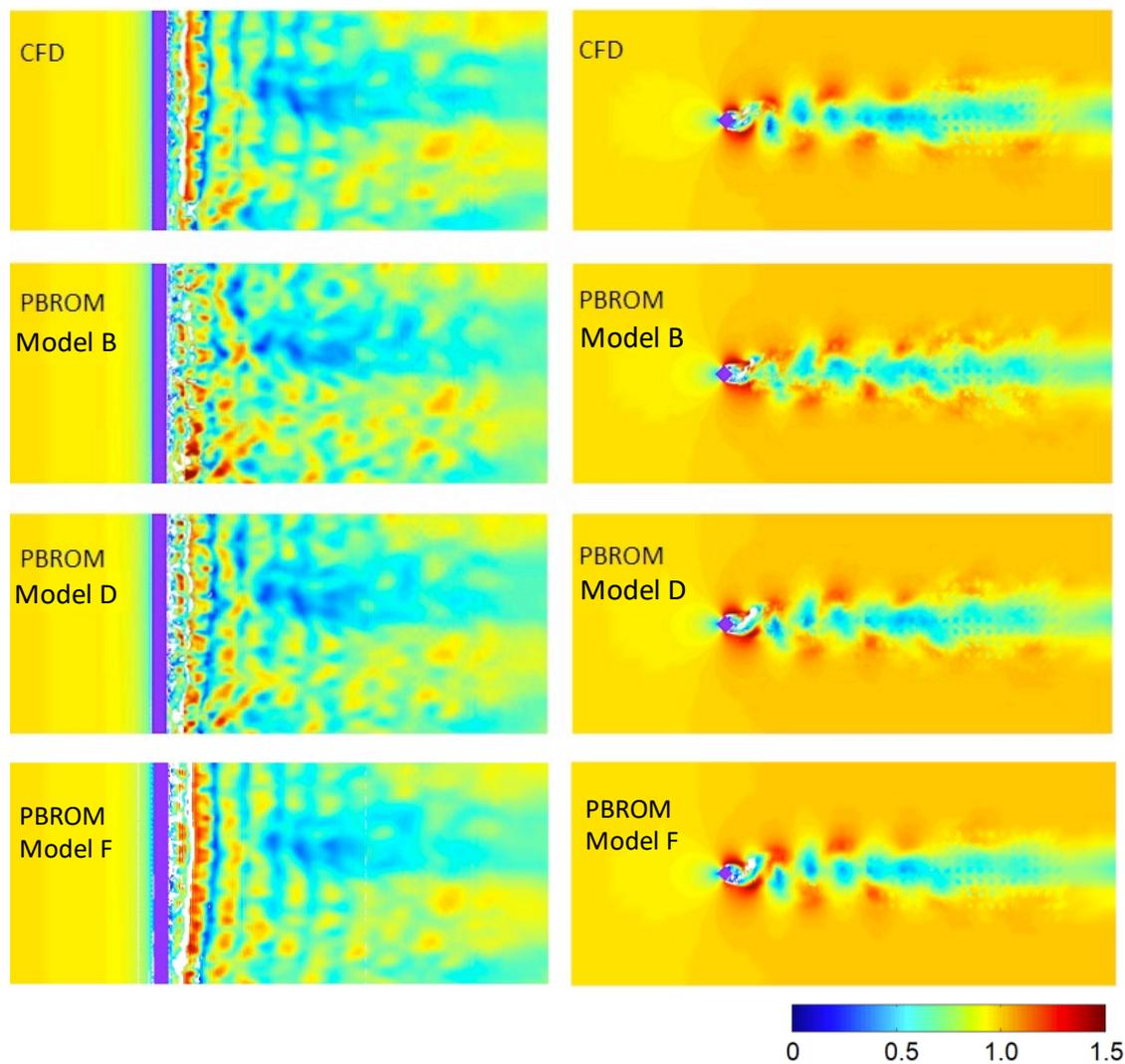

Figure 18. Velocity magnitude (normalized) on vertical (left) & horizontal (right) centre planes at $t = 5s$.

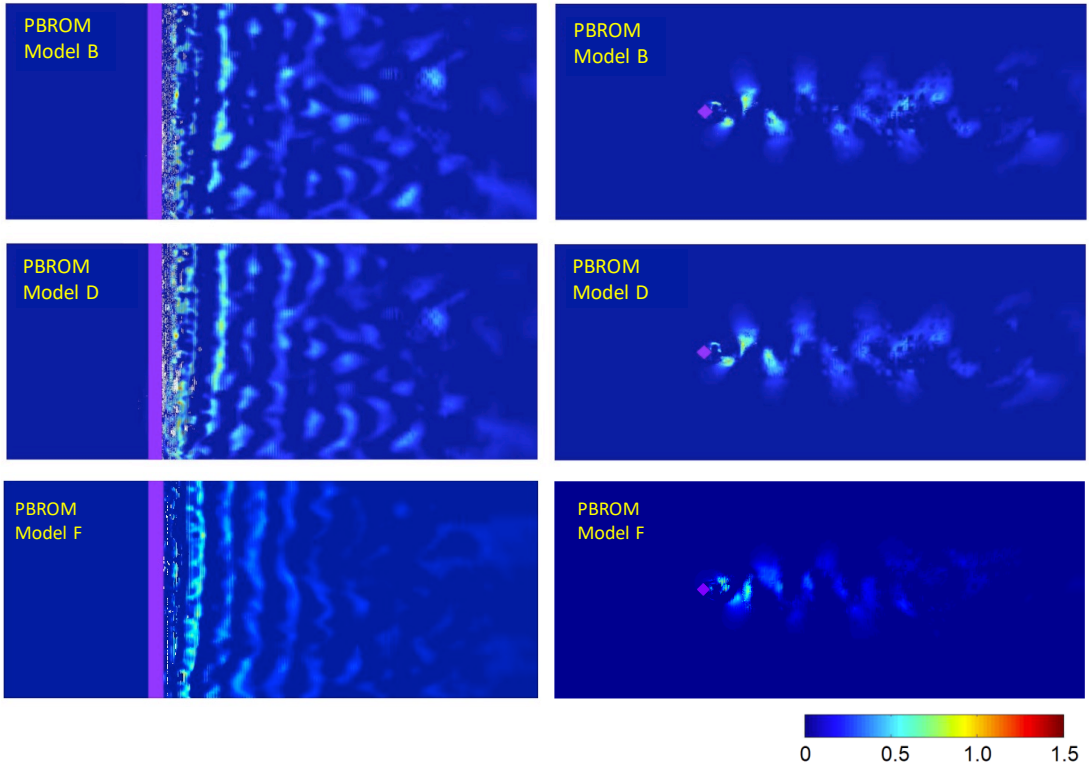

Figure 19. Velocity magnitude differences from CFD (normalized) on vertical (left) & horizontal (right) centre planes at $t = 5s$.

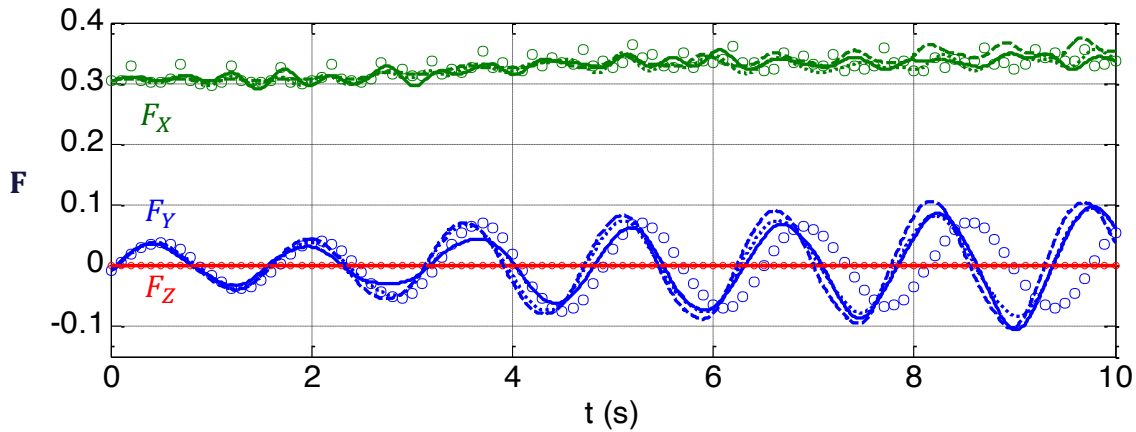

Figure 20. Time-series of the force components on the cylinder obtained from model B (thin dash lines), model D (dot lines), model F (dash lines) and CFD (symbols).

With the improvements by using residual closures, models D and F predict the forces on the cylinder more accurate in term of magnitudes in the first 5s, as shown in Figure 20,. The PBROM model B under-predicts the magnitude of transverse force, i.e. $F_Y$, during the first 5s and starts to over-predict after that. As the result of frequency shifting in of predicted velocity and presure modes, all models produces higher frequency of $F_Y$.

### 3.6 Discussion on the PBROM models

To model complex flow fields such as turbulent flows by a PBROM model, it is obvious that a large number of modes should be used in order to capture important flow features. However, using more modes will increase model size and hence reduce model efficiency. Furthermore, low energy modes and the interactions of these modes with the high energy modes in PBROM simulations are important for energy transfer among the modes and energy dissipation. Increase the number of modes without proper modelling of turbulent viscosity for the dissipation of energy, and the interactions among the modes (including the discarded modes) and modelling the transfer of energy may lead to excessive growth in amplitudes of some modes while decay in other modes.

In the present test cases, a space-varying turbulent viscosity model is shown working reasonably well. This turbulent viscosity field can be directly obtained from the temporal averaging of an ensemble of pre-simulated snapshots. For a general turbulent flow, the temporal averaged field may not adequately represent the dynamics of the turbulent viscosity. In that case, inclusion of the time-varying components of turbulent viscosity may be considered although this will increase the complexity of the PBROM model due to additional models for their modal coefficients. In the present study, these models have not been explored.

This study also shows the effects of discarded low energy modes and interactions among the modes to the modelling turbulence flows. These effects are modelled very well by an ELM and a NARX closures in 2D cases. The remarkable improvements are the growing anmplitues and frequency shifts of major high energy modes being handled well. The growing amplitude is handled well in the 3D case but much less improvement in frequency shift is achieved. This is probably due to more complex interactions of horizontal velocity modes with the vertical velocity modes which the current closure models fail to capture. Further study on these closures or other ML methods to better resolve these interactions should be carried out.

In these test cases, the computational complexity of the training of the ELM closures is the pseudo-inverse of a matrix having a dimension of the number of hidden layers multiplying with the number of basis vectors. This operation is of an order cheaper than that of the PBROM model and can be performed online. The training of the NARX closures is relatively longer, approximately at the same order of the PBROM run-time. The computational complexity of the closures involving only several matrix vector multiplications, which is also at an order lower than the computational complexity of the PBROM model does not affect the overall efficiency of the combined model. The CPU-times in Table 2 show enormous speedup gains of the reduced order model for these cases. From 2D to 3D cases, the degree of freedom of the CFD simulation increases drastically, while that of PBROM often increases by a few times. The speedup is therefore much more for larger size problems.

Table 2 Speed-up of PBROM models compared to full CFD to simulate 1s of model time.

| CFD size (million cells) | PBROM size (no. of modes) | CFD CPU-time (s) | PBROM CPU-time (s) | Speedup |
|---|---|---|---|---|
| 0.125 (2D) | 20 | 13,500 | 1.5 | 8,950 |
| 4.2 (3D) | 60 | 129,600 | 6.7 | 19,340 |

## 4   Conclusions

The paper presents a Projection-Based Reduced-Order Model for simulations of turbulent flows. As often stated in PBROM simulations, a large number of basis vectors should be used to capture important flow features. The presented results show that, due to strong nonlinearities and energy cascade and dissipation in the high Re turbulent flows, without proper modelling of turbulent viscosity and interactions among the modes may lead to instabilities in the model prediction. Several models for turbulent viscosity including a space-time-average, a space-varying and a space-time-varying models for turbulent viscosity are presented. Good prediction accuracies are achieved with the space- and time-space-varying models. Two residual closures based on the Extreme Learning Machine and the Nonlinear AutoRegressive eXogenous methods are employed to model effects of discarded low energy modes and interactions among the modes in the projection of momentum equations. The closures are demonstrated capable of recovering the information lost due to mode truncation in PBROM while not affecting the efficiency of the overall model. Though only ELM and NARX are implemented in this study and demonstrate promising results, other ML methods should also be explored for further improvements and for a wider applicability of the residual closure approach.